\documentclass[10pt,conference]{IEEEtran}
\IEEEoverridecommandlockouts
\usepackage{geometry}
\usepackage{tikz}

\newgeometry{top=0.75in, bottom=1in, left=0.625in, right=0.625in}

\usepackage{cite}
\usepackage{amsmath,amssymb,amsfonts}
\usepackage{graphicx}
\usepackage{textcomp}
\usepackage{xcolor}
\def\BibTeX{{\rm B\kern-.05em{\sc i\kern-.025em b}\kern-.08em
    T\kern-.1667em\lower.7ex\hbox{E}\kern-.125emX}}

\usepackage{algorithmicx}
\usepackage{algorithm}
\usepackage[bookmarks=false]{hyperref}
\usepackage{multirow}
\usepackage{subcaption}

\def\NoNumber#1{{\def\alglinenumber##1{}\State #1}\addtocounter{ALG@line}{-1}}

\newcommand\copyrighttext{%
  \footnotesize This article has been accepted for presentation in \href{https://globecom2023.ieee-globecom.org/}{IEEE GLOBECOM 2023}, but has not been fully edited. Content may change prior to final publication. \textcopyright 2023 IEEE. Personal use of this material is permitted. Permission from IEEE must be obtained for all other uses, in any current or future media, including reprinting/republishing this material for advertising or promotional purposes, creating new collective works, for resale or redistribution to servers or lists, or reuse of any copyrighted component of this work in other works.}
\newcommand\copyrightnotice{%
\begin{tikzpicture}[remember picture,overlay]
\node[anchor=north,yshift=-5pt] at (current page.north) {\fbox{\parbox{\dimexpr\textwidth-\fboxsep-\fboxrule\relax}{\copyrighttext}}};
\end{tikzpicture}%
}

\begin{document}

\title{How Does Forecasting Affect the Convergence of DRL Techniques in O-RAN Slicing?
}

\author{\IEEEauthorblockN{Ahmad M. Nagib\IEEEauthorrefmark{1}\IEEEauthorrefmark{3},
Hatem Abou-zeid\IEEEauthorrefmark{2},
Hossam S. Hassanein\IEEEauthorrefmark{1}
}
\IEEEauthorblockA{\IEEEauthorrefmark{1}\textit{School of Computing}, 
\textit{Queen's University}, Canada, \{ahmad, hossam\}@cs.queensu.ca \\}
\IEEEauthorblockA{\IEEEauthorrefmark{2}\textit{Department of Electrical and Software Engineering}, 
\textit{University of Calgary},
 Canada, hatem.abouzeid@ucalgary.ca \\
}
\IEEEauthorblockA{\IEEEauthorrefmark{3}\textit{Faculty of Computers and Artificial Intelligence}, 
\textit{Cairo University},
 Egypt \\
}
\\[-6.7ex]
\thanks{ This research was supported by the Natural Sciences and Engineering Research Council of Canada (NSERC) under Grant RGPIN-2019-05667 and Grant RGPIN-2021-04050.}
}

\maketitle
\copyrightnotice

\begin{abstract}

The success of immersive applications such as virtual reality (VR) gaming and metaverse services depends on low latency and reliable connectivity. To provide seamless user experiences, the open radio access network (O-RAN) architecture and 6G networks are expected to play a crucial role. RAN slicing, a critical component of the O-RAN paradigm, enables network resources to be allocated based on the needs of immersive services, creating multiple virtual networks on a single physical infrastructure. In the O-RAN literature, deep reinforcement learning (DRL) algorithms are commonly used to optimize resource allocation. However, the practical adoption of DRL in live deployments has been sluggish. This is primarily due to the slow convergence and performance instabilities suffered by the DRL agents both upon initial deployment and when there are significant changes in network conditions. In this paper, we investigate the impact of time series forecasting of traffic demands on the convergence of the DRL-based slicing agents. For that, we conduct an exhaustive experiment that supports multiple services including real VR gaming traffic. We then propose a novel \emph{forecasting-aided} DRL approach and its respective O-RAN practical deployment workflow to enhance DRL convergence. Our approach shows up to 22.8\%, 86.3\%, and 300\% improvements in the average initial reward value, convergence rate, and number of converged scenarios respectively, enhancing the generalizability of the DRL agents compared with the implemented baselines. The results also indicate that our approach is robust against forecasting errors and that forecasting models do not have to be ideal.

\end{abstract}


\section{Introduction}

The open radio access network (O-RAN) paradigm and 6G networks are expected to play a crucial role in making immersive applications a reality \cite{10054381}. Services such as virtual reality (VR) gaming and metaverse applications require low latency and reliable connectivity to provide a seamless and immersive user experience. Radio access network (RAN) slicing, a critical component of the O-RAN architecture, enables the creation of multiple virtual RANs on a single physical infrastructure. This allows for improved user experiences ensuring that users have the necessary resources to engage in immersive activities \cite{10065486}.

O-RAN enables mobile network operators (MNOs) to deploy their own applications (xApps) to intelligently control the various network functionalities in near-real-time (near-RT) via standard open interfaces \cite{9627832}. Deep reinforcement learning (DRL) algorithms are among the promising tools used to design O-RAN-compliant data-driven xApps \cite{9812489}. Although O-RAN intelligent controllers offer promising advantages \cite{9627832}, the practical adoption of DRL algorithms in live deployments has been sluggish \cite{9812489}. This is primarily because of the slow convergence and performance instability suffered by DRL agents \cite{9903386}. This becomes apparent when agents are newly deployed in a live network or experience substantial changes in network conditions \cite{10075524}. DRL convergence to the optimal RAN configuration needs to be quick and stable so that the users' quality of experience (QoE) is not affected. Nonetheless, due to the stochastic nature of 6G systems and the exploratory behavior of DRL agents, it may require thousands of time steps to regain stability. This holds significant importance in O-RAN deployments, as 6G networks can only afford a few exploration iterations while optimizing near-RT O-RAN functionalities \cite{9903386}.

Traffic demand forecasting \cite{10.1145/3323679.3326521, 9500858} is a powerful tool that can be utilized to enhance the performance of O-RAN slicing intelligent controllers. Since 6G networks are expected to support a wide range of immersive services, accurate forecasting can help MNOs proactively optimize the allocation of network resources to the admitted slices. This ensures that each slice has sufficient capacity to meet its service level agreements (SLAs). Combining traffic demand forecasting with a flexible tool such as DRL can enhance the convergence of DRL-based slicing and its generalizability. This enables the DRL agent to make informed slicing decisions based on the current network conditions while also considering the forecasted conditions.

In this paper, we investigate ways to leverage the power of time series forecasting for more robust and generalizable DRL-based O-RAN slicing. Moreover, we propose an O-RAN intelligent deployment workflow that incorporates a forecasting module to enhance convergence. The contribution of this research study can be summarized as follows:

\begin{itemize}

\item We propose a novel \emph{forecasting-aided} algorithm to enhance the convergence and generalizability of O-RAN slicing DRL agents. A forecasting model is employed to predict the future contribution of slices to the overall traffic demand and a resource allocation configuration is suggested accordingly. This acts as a guide for the DRL agent when allocating resources to slices. Hence, the agent considers both its policy and the forecasted demand levels when taking allocation action given a certain situation.

\item We propose an O-RAN deployment workflow that incorporates our \emph{forecasting-aided} approach in the O-RAN architecture to guide the convergence of DRL-based xApps.

\item We conduct an exhaustive performance study that supports multiple services including live VR gaming data to examine the impact of forecasting and its errors on the convergence of DRL-based O-RAN slicing. We then compare our approach against three implemented baselines. Our approach shows up to 22.8\%, 86.3\%, and 300\% improvements in the average initial reward value, convergence rate, and number of converged scenarios respectively. The results also demonstrate our approach's robustness against forecast errors that follow a Gaussian distribution with a standard deviation up to 0.25, given that the range of the forecasted values is 1. The implementations of the proposed approach and baselines are publicly available\footnote{{Available at \href{http://www.github.com/ahmadnagib/forecasting-aided-DRL}{http://www.github.com/ahmadnagib/forecasting-aided-DRL}}} to facilitate research on trustworthy DRL in O-RAN.

\end{itemize}

To the best of our knowledge, this is the first study to 1) identify the need, and investigate the effect of time series forecasting on the convergence of DRL-based O-RAN slicing, especially for immersive 6G applications, and 2) propose an algorithm to improve DRL convergence by using a novel form of \emph{forecasting-aided} DRL.

The paper's remaining sections are structured as follows: Section \ref{sec:related} presents a discussion on related work. Section \ref{sec:proposal} details the \emph{forecasting-aided} DRL approach, O-RAN workflow, and the baselines proposed in this study. In Section \ref{sec:results}, a description of the experimental setup, and an analysis of the results are provided. Finally, Section \ref{sec:conclusion} concludes our work and presents potential future directions.

\section{Related Work}
\label{sec:related}

The challenge of slow DRL convergence has been recently addressed using approaches such as transfer learning, meta-learning, structure awareness, and heuristics \cite{9903386}. However, the focus of this paper is exploring the effect of time series forecasting on the convergence of DRL-based O-RAN slicing. Several research studies have explored the use of forecasting in resource allocation and network slicing. Nonetheless, most of these studies use statistical and machine learning (ML)-based forecasting approaches to optimize slicing directly \cite{SSENGONZI2022100142}. For instance, the work in \cite{10.1145/3323679.3326521} extends long short-term memory (LSTM) neural networks to forecast the physical resource block (PRB) utilization. Consequently, the PRB allocation to slices is made based on such a forecast.

Only a few studies make use of forecasting models to enhance DRL. The authors of \cite{9057490} propose a traffic offloading scheme that combines deep Q-network and traffic demand forecasting. A forecasting model uses the raw data collected from the DRL environment to predict traffic load statistics as a representation of the DRL state. Then, the DRL agent makes offloading decisions according to such a state. The results show that this approach outperforms tabular Q-learning. Similarly, the authors of \cite{9340607} use a forecasting model to predict the mobile traffic volume. Hence, such a forecasted value is used as part of the state of a base station (BS) sleep control DRL agent.

Both studies, however, do not investigate the impact of forecasting on the convergence performance of the used DRL algorithms. Furthermore, the surveyed studies only utilize the forecasting model as part of a pre-processing step for state representation. Finally, the live network deployment, especially in the context of O-RAN, has not been addressed.

\section{Forecasting-aided DRL-based O-RAN Slicing}
\label{sec:proposal}

We propose to utilize a forecasting module to guide the DRL agent when newly deployed in a live network or when the network conditions change significantly as described in Section \ref{sec:flow}. This allows the agent to consider future traffic demand when allocating resources to the available slices with the goal of meeting slices' SLAs. For that, we also propose a deployment workflow to enhance the DRL convergence and generalizability in the context of the O-RAN architecture.

\begin{figure*}[ht]
\centering
\includegraphics[width=0.95\linewidth]{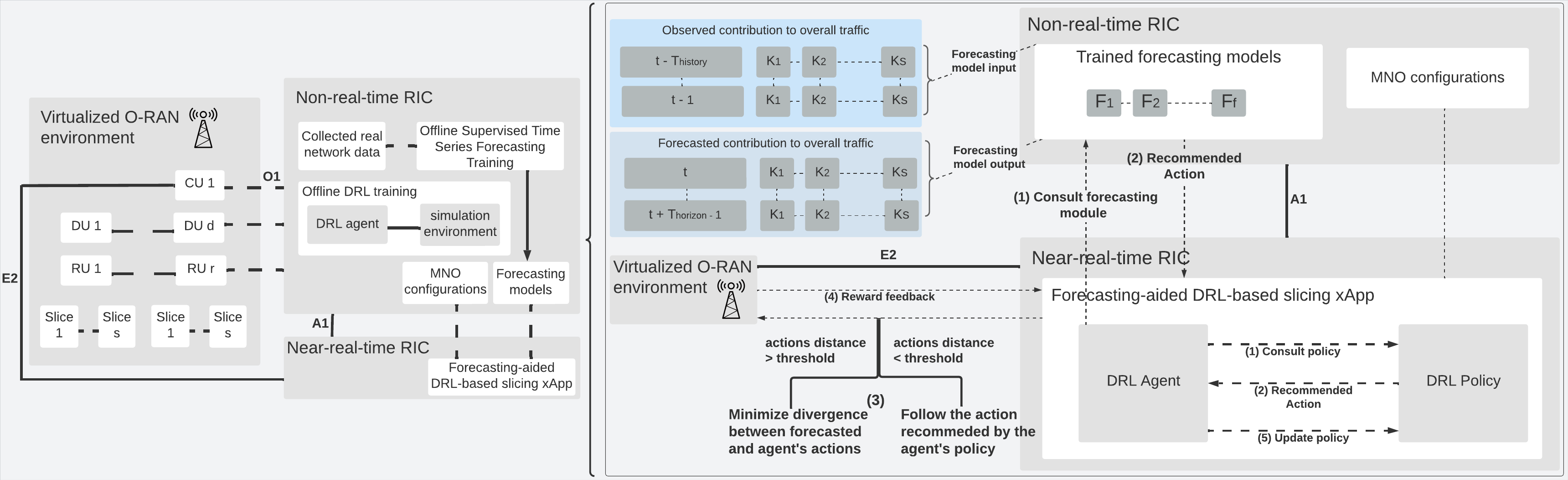}
\vspace{-1ex}
 \setlength{\belowcaptionskip}{-12pt} 
\caption{Proposed forecasting-aided DRL-based O-RAN slicing system.}
\label{fig:oran_arch}
\end{figure*}

\subsection{System Model}
\label{model}

In this paper, we are concerned with the downlink case of the radio access part of O-RAN slicing. Radio resource allocation in slicing aims at assigning the limited available PRBs to the admitted slices while satisfying the slices' various requirements. The problem can be formulated as follows \cite{10075524}:

A BS supports a range of services realized through a set of virtual slices, $\mathcal{S}=\{1,2, \ldots, S\}$. Such slices share the available bandwidth, $B$. Each BS has a set of user equipments (UEs), $\mathcal{U}=\{1,2, \ldots, U\}$, connected to it. A UE, denoted as $u$, is capable of requesting a single service type for downlink transmission at any given moment. Users associated with a particular slice, $s$, generate a set of requests, $\mathcal{R}_s=\left\{1,2, \ldots, R_s\right\}$. The overall demand, $D_s$, of these users can be denoted as follows:
\begin{equation}
D_s=\sum_{r_s \in R_s} d_{r_s},
\end{equation}

where $d_{r_s}$ is the demand of a request, $r_s$, made by a user associated with slice $s$. Furthermore, the contribution of a slice, $s$, to the total BS's traffic demand at a slicing step, $t$, is:
\begin{equation}
\label{contribution}
\kappa_s(t)=\frac{D_s(t)}{\sum_{i=1}^{\|S\|} D_i(t)}
\end{equation}

PRB allocation among the available slices, $S$, can be represented by the vector, ${a \in \rm I\!R ^S}$. At the start of a slicing window, a RAN slicing controller selects a slicing PRB allocation configuration, ${a}$, out of the $A$ feasible configurations, where $\mathcal{A}=\{1,2, \ldots, A\}$. Consequently, the system performance, represented in terms of the latency of the admitted slices within the context of this paper, is impacted. This is primarily influenced by a queue maintained at the BS.

\subsection{Reinforcement Learning Mapping}
\label{sec:mapping}

For the traditional DRL approach, we follow the mapping in Table \ref{tab:drl_parameters}. The system state is defined as the slices' contribution to the total BS's traffic within the past slicing window, that is,
\vspace{-1.25ex}
\begin{equation}
\label{eq:state}
\kappa = (\kappa_{1}(t-1), . . . , \kappa_s(t-1), . . . ,\kappa_{S}(t-1)) 
\end{equation}
\vspace{-3ex}

The DRL agent observes such a state and takes action accordingly at the start of each slicing step to decide the PRB allocation for each slice, that is,
\vspace{-1.25ex}
\begin{equation}
\label{eq:action}
a = (b_{1}, . . . , b_{s}, . . . ,b_{S}), 
\text { subject to } b_{1} + . . . + b_{S} = B 
\end{equation}
\vspace{-3.5ex}

The reward function is used to optimize the allocation process. We use sigmoid function-based rewards similar to the one proposed in \cite{10075524}. This enables controlling the effect of approaching the performance threshold defined by each slice's SLAs. In this study, we prioritize immersive services, and hence, their latency requirements. Therefore, the reward function is a weighted sum of an inverse form of latency and can be defined as follows:
\vspace{-1ex}
\begin{equation}
\label{equation:reward}
\begin{aligned}
 R = \sum_{s=1}^{\|S\|} w_{s}\ *\ \frac{\mathrm{1} }{\mathrm{1} + e^{\ c1_{s}\ *\ (\ {l}_{s}\ -\ c2_{s}\ )} }
\end{aligned}
\end{equation}
\vspace{-2ex}

where $l_{s}$ is the average latency underwent by slice $s$'s UEs during the preceding slicing window, at $t-1$. The weight, $w_{s}$, defines the priority of satisfying the delay requirement of slice $s$. $c1$ configures the slope of the sigmoid function, and hence, controls the starting point for penalizing the agent's actions. Moreover, $c2$ defines the inflection point that reflects the latency performance threshold for each slice based on its SLAs.

\subsection{Proposed Forecasting-aided O-RAN Architecture}
\label{sec:flow}

We propose a deployment workflow that incorporates a forecasting module as part of the O-RAN architecture to guide DRL-based xApps. This is primarily needed upon the initial xApp deployment and when there are significant changes in network conditions. This module includes forecasting models that predict relevant network conditions in the future such as the traffic demand of various slices. Consequently, the module interacts with the DRL agents via O-RAN's A1 interface to guide and enhance their convergence performance as described in the next subsections. Fig. \ref{fig:oran_arch} shows the overall system architecture (left) and the interaction steps between the DRL agent, the forecasting module, and the virtualized O-RAN environment in the proposed approach (right).

As seen in the figure, the traffic demand forecasting is based on observations made during a time window, $T_{\mathrm{history}}, [t-T_{\mathrm{\text{history}}}, \left.t-1\right]$. The contribution of slice $s$ to the overall traffic demand in such a period can be described by the following vector:
\vspace{-3ex}

\begin{equation}
    \begin{split}
\mathbf{\kappa}^{(\mathbf{s})}(t) = \Bigl(\kappa^{(\mathbf{s})}(t-T_{\mathrm{\text{history}}}), \kappa^{(\mathbf{s})}(t-(T_{\mathrm{\text{history}}}-1)), \\
\cdots, \kappa^{(\mathbf{s})}(t-1)\Bigl)
    \end{split}
\end{equation}

\vspace{-1.25ex}

Based on such an observed history of traffic demand, the forecasting model, $\mathbb{F}$, provides the predicted traffic demand for the slicing window that is about to begin or a longer time window, $T_{\mathrm{\text{horizon}}}$, for the time period $[t$, $\left.t+T_{\text{horizon}}-1\right]$. The demand in this period can be denoted as follows:

\vspace{-3ex}
\begin{equation}
\label{eqn:forecasted_demand}
    \begin{split}
\hat{\mathbf{\kappa}}^{(\mathbf{s})}=\Bigl(\hat{\kappa}^{(s)}(t), \hat{\kappa}^{(s)}(t+1), \text{ } \cdots, \hat{\kappa}^{(s)}(t+T_{\text {horizon}}-1)\Bigl)
    \end{split}
\end{equation}
\vspace{-2ex}

Accordingly, the forecasting model depicted in Fig. \ref{fig:oran_arch} predicts the slices' future contribution to the overall BS traffic demand as follows:

\begin{equation}
\label{equation:forecast}
\begin{aligned}
& \mathbb{F}: {\mathbb{R}}^{T_{\mathrm{\text{history}}}*S} \rightarrow \mathbb{R}^{T_{\text {horizon}}*S} \\
& \mathbf{\kappa}[t-T_{\mathrm{\text{history}}}, \left.t-1\right] \rightarrow \hat{\mathbf{\kappa}}[t, \left.t+T_{\text{horizon}}-1\right]
\end{aligned}
\end{equation}

where $\hat{\mathbf{\kappa}}$ denotes the slices' forecasted contribution to the overall traffic demand. For the purpose of this paper, we imitate the behavior of forecasting models with errors that follow Gaussian distributions as detailed in Section \ref{sec:results}. Hence, forecasting model training is not addressed.

\begin{table*}
\centering
\caption{Experiment Setup: RAN Slicing DRL Design}
\begin{tabular}{|p{1in}|p{1in}|p{1in}|p{1in}|}
\hline
\multicolumn{1}{|l|}{\textbf{State}}                                                                                                                        & \multicolumn{3}{l|}{\begin{tabular}[c]{@{}l@{}}Slices' contribution to the overall BS's traffic within a specific time window as defined in (\ref{contribution})\end{tabular}}                                                     \\ \hline
\multicolumn{1}{|l|}{\textbf{Action}}                                                                                                                       & \multicolumn{3}{l|}{\begin{tabular}[c]{@{}l@{}}PRBs allocated to each slice as defined in (\ref{eq:action})\end{tabular}} \\ \hline
\multicolumn{1}{|l|}{\textbf{Reward function}}                                                                                                                       & \multicolumn{3}{l|}{\begin{tabular}[c]{@{}l@{}}A weighted sum of a sigmoid function of the average latency experienced in a slicing window by the \\various slices as defined in (\ref{equation:reward})\end{tabular}} \\ \hline
\multicolumn{1}{|l|}{\textbf{Reward function weights}}                                                                                                                       & \multicolumn{3}{l|}{VoNR: 0.1, VR gaming: 0.7, Video: 0.2}                                                                                                                                                                \\ \hline
\multicolumn{1}{|l|}{\textbf{DRL algorithm}}                                                                                                                       & \multicolumn{3}{l|}{Proximal Policy Optimization (PPO)}                                                                                                                                                                \\ \hline
\multicolumn{1}{|l|}{\textbf{Learning steps per run}}                                                                                                                       & \multicolumn{3}{l|}{10,000}                                                                                                                                                                \\ \hline
\multicolumn{1}{|l|}{\textbf{Exploration rate}}                                                                                                                       & \multicolumn{3}{l|}{0.5}                                                                                                                                                                \\ \hline
\multicolumn{1}{|l|}{\textbf{Exploration decay rate}}                                                                                                                       & \multicolumn{3}{l|}{0.5 (every 200 steps)}                                                                                                                                                                \\ \hline
\multicolumn{1}{|l|}{\textbf{Action distance threshold}}                                                                                                                       & \multicolumn{3}{l|}{7\%}                                                                                                                                                                \\ \hline
\multicolumn{1}{|l|}{\textbf{Learning rate}}                                                                                                                       & \multicolumn{3}{l|}{0.01}                                                                                                                                                                \\ \hline
\multicolumn{1}{|l|}{\textbf{Batch size}}                                                                                                                       & \multicolumn{3}{l|}{4}                                                                                                                                                                \\ \hline

\end{tabular}
\vspace{-3ex}
\label{tab:drl_parameters}
\end{table*}

\subsection{Proposed Forecasting-aided DRL-based O-RAN Slicing} 
\label{sec:proposed_approach}

\setlength{\textfloatsep}{4pt}

\begin{algorithm}

\caption{Proposed Forecasting-aided DRL Approach}
\label{alg:one}
\textbf{Input:} trained forecasting model, $\mathbb{F}$, traffic demand historical observations of size $T_{\text {history}}$, forecast horizon, $T_{\text {horizon}}$, current state, $\kappa$, current DRL policy, $\pi$, set of possible actions, $A$, actions distance threshold, $\gamma_{\text{threshold}}$ \\
\textbf{Output:} distilled action, $a_{\text{distilled}}$ as defined in Section \ref{sec:proposed_approach}
\vspace{-1ex}

\hrulefill

\begin{algorithmic}[1]

\State \textbf{while} $t < T$ \textbf{do}:
\State \quad Forecast $\hat{\kappa}$ for the next $T_{\text {horizon}}$ time steps, using $\mathbb{F}$
\State \quad Generate an action, $a_{\text{forecast}}$, purely based on $\hat{\kappa}$ 
\State \quad Consult $\pi$ given $\kappa$, and get the recommended action, $a_{\pi}$
\State \quad \textbf{if} $\gamma(a_{\pi}, a_{\text{forecast}}) > \gamma_{\text{threshold}}$ \textbf{do}:
    \State \quad \quad Find the midpoint between the vectors, $a_{\pi}$ and $a_{\text{forecast}}$
    \State \quad \quad Select an action, $a_{\text{distilled}}$, closest to the midpoint to \NoNumber{\quad\quad minimize the divergence between $a_{\pi}$ and $a_{\text{forecast}}$ }
    \State \quad \quad Take the distilled action, $a_{\text{distilled}}$, to allocate PRBs for\NoNumber{\quad\quad each admitted slice}
    \State \quad \quad Update the value function, $V$, based on the received\NoNumber{\quad\quad reward, $R$}
\State \quad \textbf{end if}
% \State \quad $t \leftarrow t + 1$
\State \textbf{end while}
\end{algorithmic}
\end{algorithm}

As detailed in Fig. \ref{fig:oran_arch}, a forecasting model is incorporated to predict future traffic demand. Accordingly, the forecasting module suggests a PRB allocation action purely based on its forecasted contribution to the total demand, $\hat{\mathbf{\kappa}}$, as defined in Equation \ref{eqn:forecasted_demand}. The DRL agent continuously monitors such a suggested guiding action via O-RAN’s A1 interface. The agent follows its current policy unless the action is significantly different from that suggested by the forecasting module. The forecasting module overwrites the agent's policy in such a case to prevent potentially damaging actions. A distilled action that minimizes the divergence between the agent's policy and the forecasting module's action is taken. The difference between two actions is measured in terms of the Euclidean distance between the action vectors as follows:

\vspace{-1.25ex}
\begin{equation}
\centering
\label{eqn:euclideandistance}
{ 
\gamma \left( {a_{\pi}},{a_{\text{forecast}}}\right)   = \sqrt {\sum _{s=1}^{S}  \left( {a_{\pi}}_{s}-{a_{\text{forecast}}}_{s}\right)^2 }
}
\end{equation}

where $a_{\pi}$ and ${a_{\text{forecast}}}$ are vectors of actions recommended by the DRL agent's policy and the forecasting module respectively. The agent does not follow the exact action recommended by the forecasting module as it may not explicitly consider the slices' SLAs, i.e., latency in our case. A distilled action that represents the midpoint between the two actions' vectors is taken instead. This prevents the agent from taking actions that contradict the forecasted demand. This additionally accommodates potential forecast errors. The distilled action is integrated into the DRL agent's learning process to speed up its convergence to the optimal slicing configuration. Algorithm \ref{alg:one} defines how a DRL-based slicing xApp seeks guidance from the forecasting module and updates its policy accordingly while it is active (i.e., $t < T$).

\vspace{-1ex}

\subsection{Baselines} 
\label{sec:baselines}

\vspace{-0.4ex}

\paragraph{Forecasting-based DRL state representation} We first implement a forecasting-based approach that embeds the forecasted traffic demand as part of the DRL state. This approach utilizes the forecasting module in a preprocessing step similar to \cite{9057490} and \cite{9340607}. We integrate this step in the O-RAN flow proposed in Section \ref{sec:flow} and use the forecasting model's output as an extra input feature to the DRL-based agent via O-RAN's A1 standardized interface. Hence, the forecasted traffic demand is embedded as part of the DRL state representation in addition to the traffic demand observed in the preceding slicing window as defined in (\ref{eq:state}), where ${\kappa \in \rm \mathbb{R}^{(T_{\text {horizon}}+1)*S}}$. The first baseline approach's steps are described in Algorithm \ref{alg:two}. 

\vspace{-0.5ex}

\begin{algorithm}
\caption{Forecasting-based DRL State Representation}
\label{alg:two}
\textbf{Input:} trained forecasting model, $\mathbb{F}$, traffic demand historical observations of size $T_{\text {history}}$, forecast horizon, $T_{\text {horizon}}$, current DRL policy, $\pi$, set of possible actions, $A$ \\
\textbf{Output:} action, $a$ as defined in (\ref{eq:action})
\vspace{-1ex}

\hrulefill

\vspace{-0.5ex}
\begin{algorithmic}[1]
\State \textbf{while} $t < T$ \textbf{do}:
\State \quad Forecast $\hat{\kappa}$ for the next $T_{\text {horizon}}$ time steps, using $\mathbb{F}$
\State \quad Embed the future traffic demand in the system state, \NoNumber{\quad i.e., ${\kappa \in \rm \mathbb{R}^{(T_{\text {horizon}}+1)*S}}$}
\State \quad Consult $\pi$ given $\kappa$, and get the recommended action, $a_{\pi}$
\State \quad Update the value function, $V$, based on the received\NoNumber{\quad reward, $R$}

\State \textbf{end while}

\end{algorithmic}
\end{algorithm}

\setlength{\textfloatsep}{20pt}

\paragraph{Non-forecasting-aided DRL} We also implement a traditional DRL approach that follows the same DRL mapping defined in Section \ref{sec:mapping}. This approach is not guided by the proposed forecasting module.

\paragraph{Non-DRL forecasting approach} Finally, we implement an approach that purely relies on forecasting to allocate resources. PRBs are allocated to each slice solely based on such a slice's forecasted contribution to the total demand, $\hat{\mathbf{\kappa}}$.

\begin{table*}
\centering
\caption{Experiment Setup: Simulation Parameters Settings}
\begin{tabular}{|p{1.2in}|p{2.142in}|p{1.67in}|p{1.3in}|}
\hline
\textbf{\begin{tabular}[c]{@{}l@{}} \end{tabular}}   & \textbf{Video}     & \textbf{VoNR}                & \textbf{VR gaming}                                                                                                  \\ \hline

\multicolumn{1}{|l|}{\textbf{Scheduling algorithm}}                                                                                                                        & \multicolumn{3}{l|}{\begin{tabular}[c]{@{}l@{}}Round-robin per 1 ms slot\end{tabular}}                                                     \\ \hline
\multicolumn{1}{|l|}{\textbf{Slicing window size}}                                                                                                                       & \multicolumn{3}{l|}{\begin{tabular}[c]{@{}l@{}}PRB allocation among slices every 100 scheduling time slots\end{tabular}} \\ 
\hline
\multicolumn{1}{|l|}{\textbf{Forecasting error }}                                                                                                                       & \multicolumn{3}{l|}{\begin{tabular}[c]{@{}l@{}}Gaussian distribution, mean = 0, standard deviation = 0, 0.1, 0.2, 0.23, 025, 0.3, 0.4\end{tabular}} \\ 
\hline
\multicolumn{1}{|l|}{\textbf{Forecasting horizons}}                                                                                                                       & \multicolumn{3}{l|}{\begin{tabular}[c]{@{}l@{}}$T_{\mathrm{\text{horizon}}}=\{1,2, \ldots, 10\}$\end{tabular}} \\ 

\hline
\textbf{\begin{tabular}[c]{@{}l@{}}Packet interarrival time \end{tabular}}   & Truncated Pareto (mean = 6 ms, max = 12.5 ms)      & Uniform (min = 0 ms, max = 160 ms)                                                       & Real VR gaming dataset \cite{9685808}                                                                                                     \\ \hline

\textbf{\begin{tabular}[c]{@{}l@{}}Packet size\end{tabular}}                & Truncated Pareto (mean = 100 B, max = 250 B) & Constant (40 B)                                                                    & Real VR gaming dataset \cite{9685808}                                                   \\ \hline
\textbf{\begin{tabular}[c]{@{}l@{}}Number of users\\ \end{tabular}}                & Poisson (max = 43, mean = 20)                       & Poisson (max = 104, mean = 70)                                                         & Poisson (max = 7, mean = 1)                                                          \\ \hline
\end{tabular}
\vspace{-2ex}
\label{tab:sim_parameters}
\end{table*}

\begin{figure*}[ht] 
    
  \begin{subfigure}[b]{0.5\linewidth}
    \centering
\includegraphics[width=0.93\linewidth]{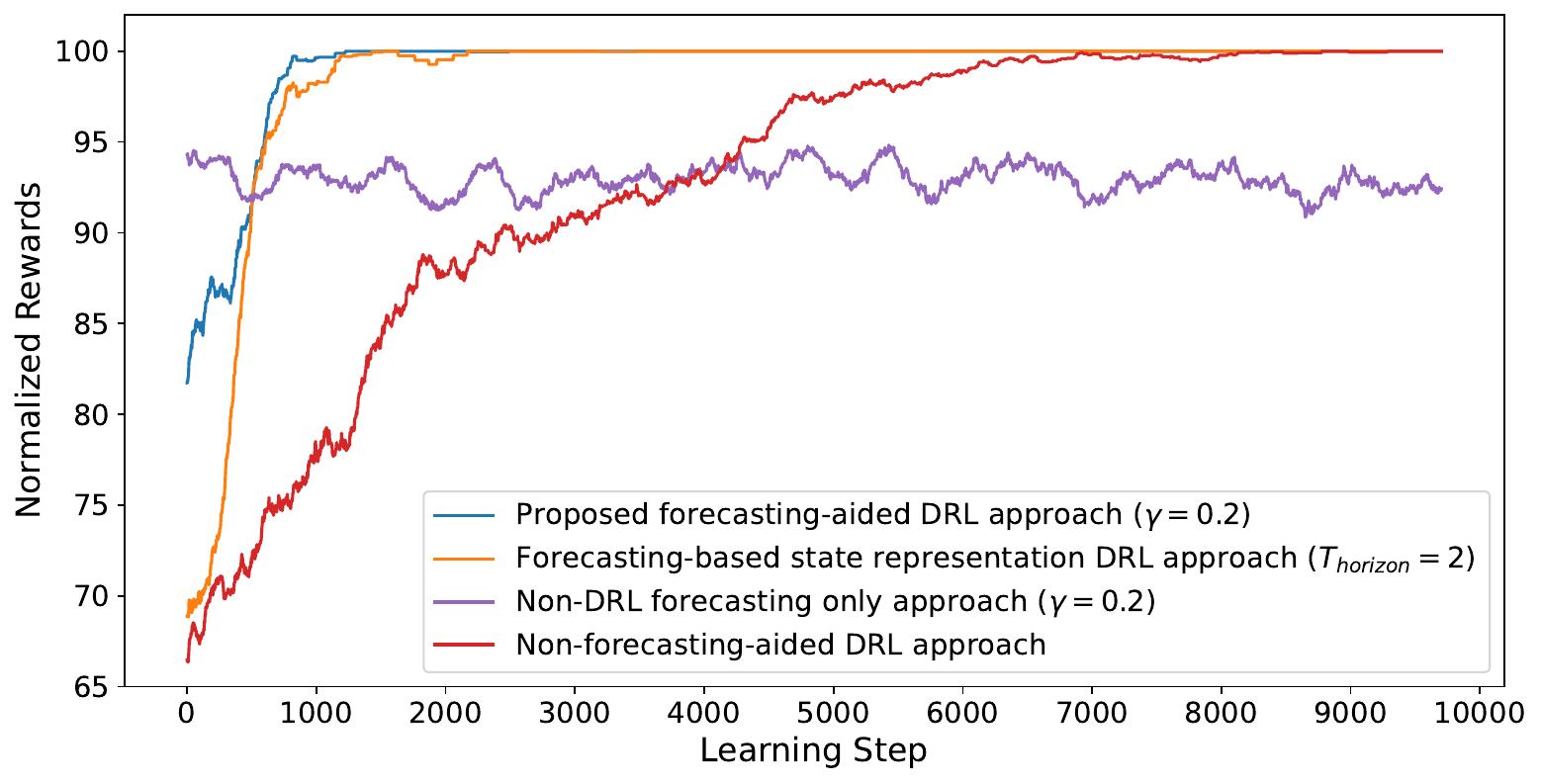} 
\vspace{-1.5ex}
    \caption{Traffic pattern 1} 
    \label{fig5:a} 
  \end{subfigure}
      \vspace{-0.25ex}
  \begin{subfigure}[b]{0.5\linewidth}
    \centering
    \includegraphics[width=0.93\linewidth]{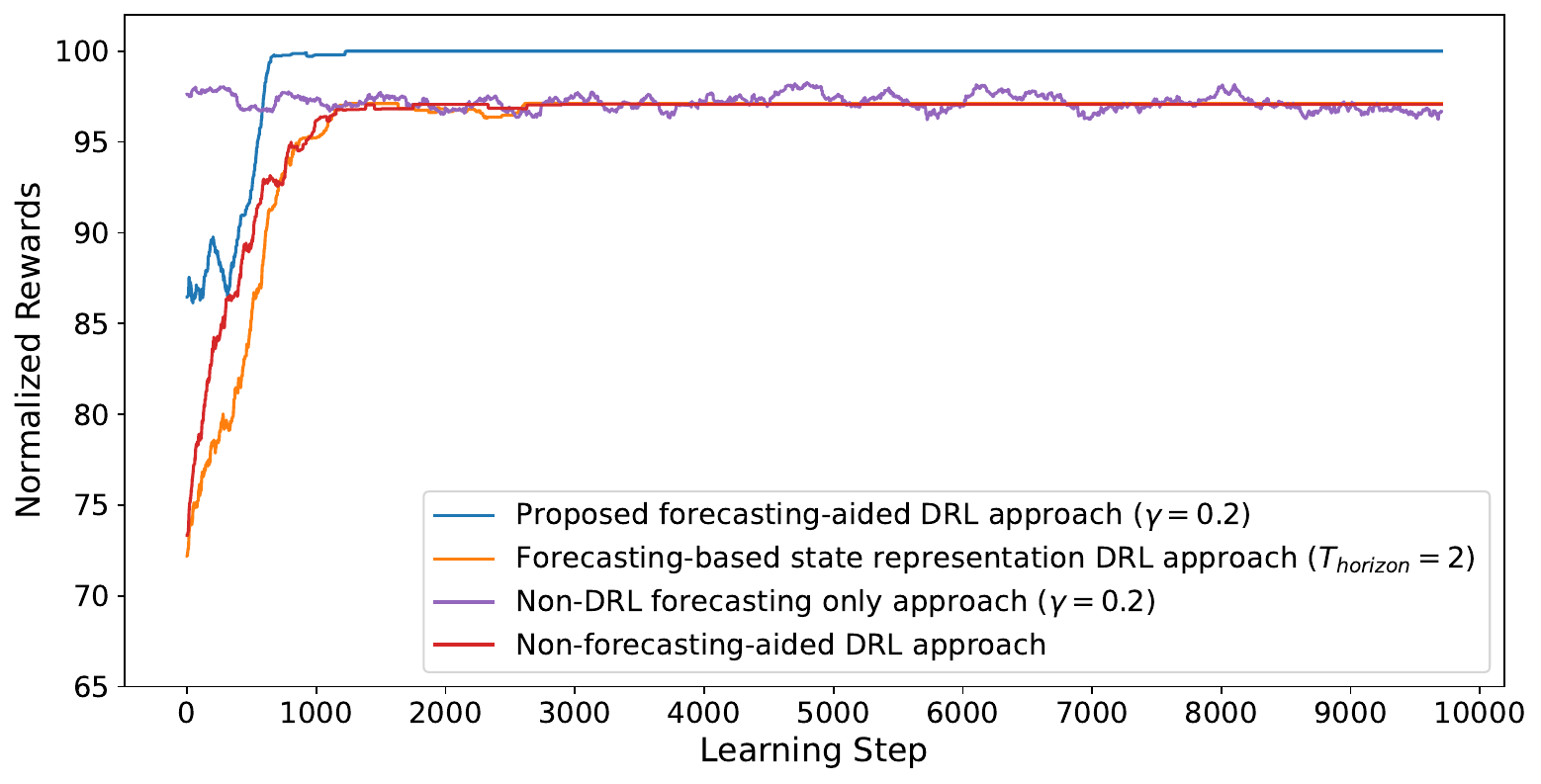} 
    \vspace{-1.5ex}
    \caption{Traffic pattern 2} 
    \label{fig5:b} 
  \end{subfigure} 
    \vspace{-0.25ex}
  \begin{subfigure}[b]{0.5\linewidth}
    \centering
    \includegraphics[width=0.93\linewidth]{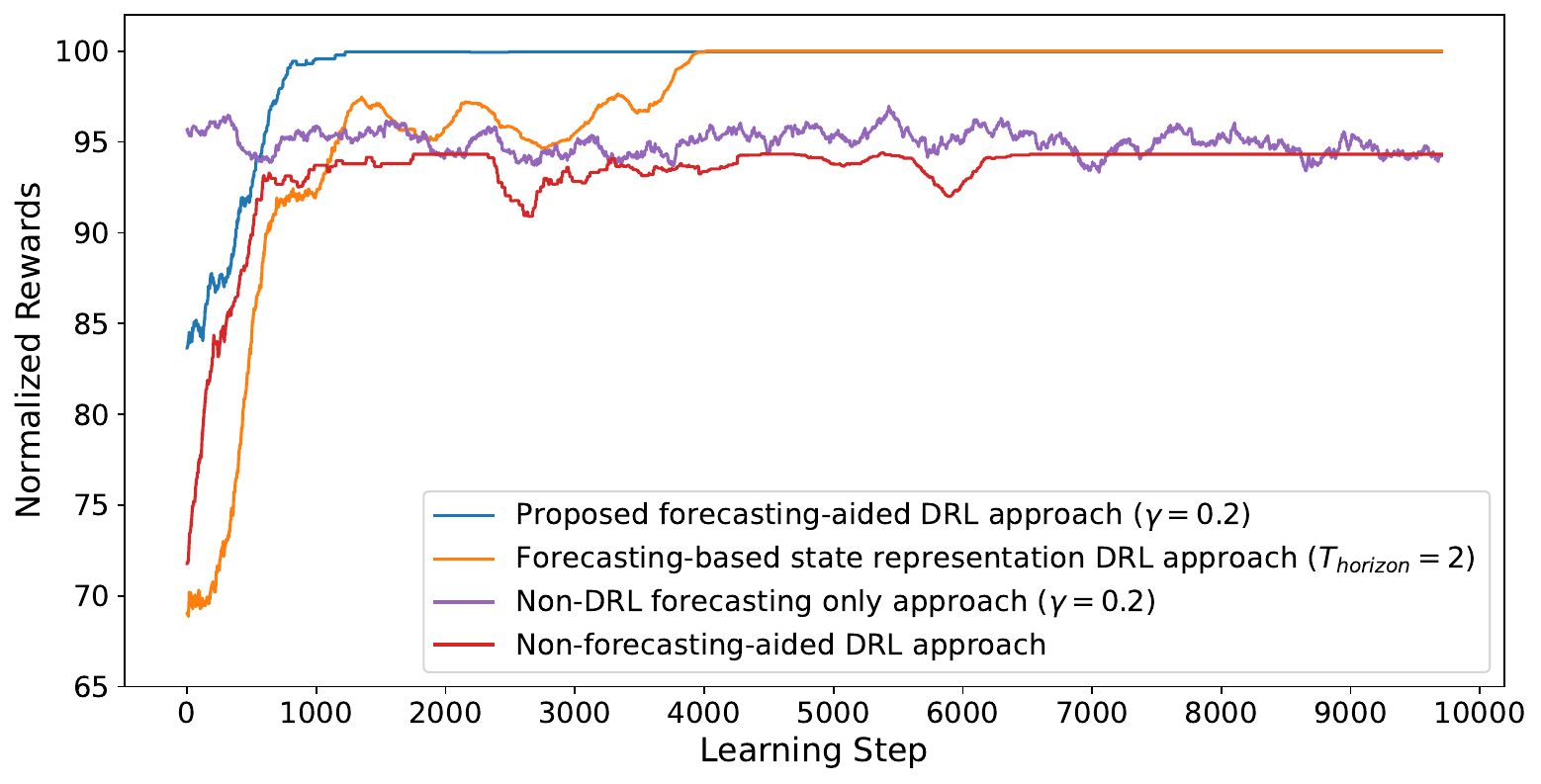} 
    \vspace{-1.5ex}
    \caption{Traffic pattern 3} 
    \label{fig5:c} 
  \end{subfigure}
    \vspace{-1ex}
  \begin{subfigure}[b]{0.5\linewidth}
    \centering
    \includegraphics[width=0.93\linewidth]{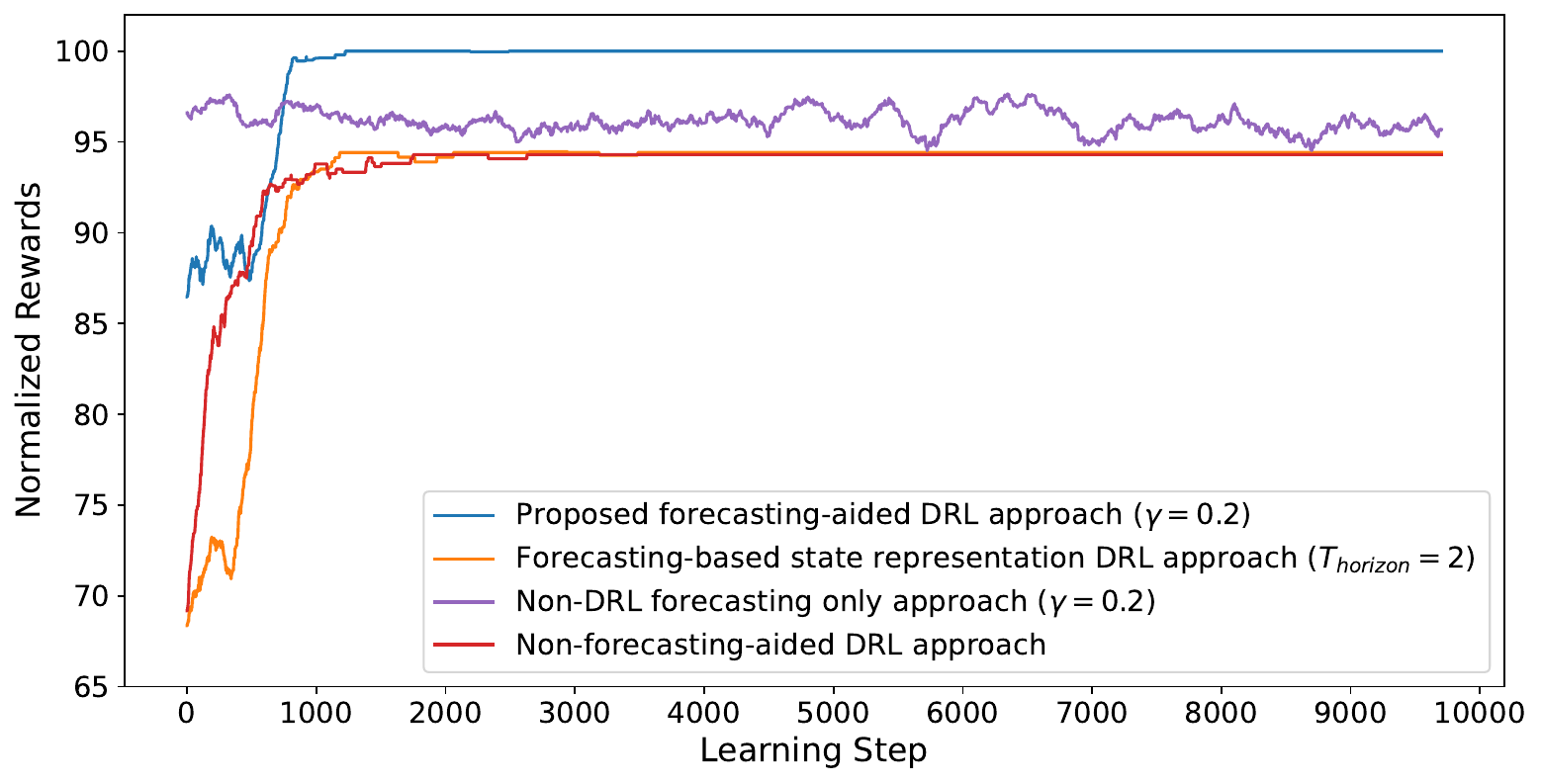} 
    \vspace{-1.5ex}
    \caption{Traffic pattern 4} 
    \label{fig5:d} 
  \end{subfigure}
      \vspace{-2ex}
     \setlength{\belowcaptionskip}{-12pt} 
  \caption{Convergence performance of the proposed forecasting-aided approach under 4 different traffic patterns.}
  \label{fig:convergence_behavior}

\end{figure*}

\section{Experiment Setup and Numerical Results}
\label{sec:results}

\subsection{Experiment Setup}

We investigate a deployment scenario using the O-RAN workflow proposed in Section \ref{sec:flow}. Hence, we restrict DRL agents' exploration as reflected by the exploration rate and its decay specified in Table \ref{tab:drl_parameters}. We conduct an exhaustive experiment that follows the mapping defined in Section \ref{sec:mapping} and implements the proposed approach. The simulation is designed to reflect extreme situations in which the available PRBs are configured to be less than the actual demand. We then compare the convergence performance of the proposed approach against the baselines defined in Section \ref{sec:baselines}. Moreover, we test the agents that follow Algorithm \ref{alg:one}, and the non-DRL forecasting approach against various forecasting errors as defined in Table \ref{tab:sim_parameters}. This allowed us to examine the effect of forecasting errors on the performance of the two approaches.

We use live VR gaming data from \cite{9685808} as an example of realistic patterns of immersive services in 6G networks. We specifically incorporate 4 different trace files reflecting traffic patterns from multiple games and distinct configurations per game. Moreover, we combine such patterns with video and voice over new radio (VoNR) traffic requests to reflect 3 different slice types in our experiment. Such requests are generated following the models defined in Table \ref{tab:sim_parameters} as in \cite{10075524}. In such models, VoNR users produce requests of small and static sizes, while VR gaming users generate the largest requests. Besides, video users experience more frequent requests than the other two services. Different constant values of $c1$ and $c2$ parameters are utilized for the slices in the reward function based on their respective latency requirements.

\vspace{-1ex}

\subsection{Numerical Results}

\vspace{-1ex}

\paragraph{Convergence Performance}

Despite the aforementioned restricted exploration settings, the proposed approach converges to the optimal allocation configurations given the scenarios shown in Fig. \ref{fig:convergence_behavior}. On the other hand, due to such restrictions, the non-forecasting-aided DRL approach fails to converge in almost all the scenarios. Moreover, the number of steps needed by our approach to converge to the optimal configuration is significantly less than the non-forecasting-aided DRL approach when both converge as in Fig. \ref{fig5:a}. This is primarily due to guidance from the forecasting module that prevents the agent from exploring potentially damaging actions. Algorithm \ref{alg:two} only includes the forecasted demand in its state so it does not directly overwrite potentially damaging actions or accommodate forecasting errors. Hence, its performance is also inferior to Algorithm \ref{alg:one}.

The figure also shows a remarkable improvement in the initial reward values of our approach compared to the other two DRL-based baselines. In our approach, a distilled action is only triggered when there is a big gap between the agent's action and that recommended by the forecasting model. Hence, only potentially damaging actions are replaced by an action closer to the forecasted conditions allowing for a safer exploration. Eventually, based on the configured action distance threshold of our approach, the forecasting model was only consulted 8.9\% of the time on average. Consequently, the DRL agent recovers quickly to a near-optimal slicing configuration and the received reward becomes relatively higher.

Relying on forecasting solely never leads to convergence given imperfect predictors as Fig. \ref{fig:convergence_behavior} suggests. This confirms our hypothesis in Section \ref{sec:proposed_approach}. The non-DRL forecasting approach does not explicitly consider the various slices' SLA fulfillment. Hence, our approach outperforms the non-DRL forecasting approach as it aims at satisfying latency requirements reflected in the reward function. 

\begin{figure}
\centering
\includegraphics[width=0.93\linewidth]{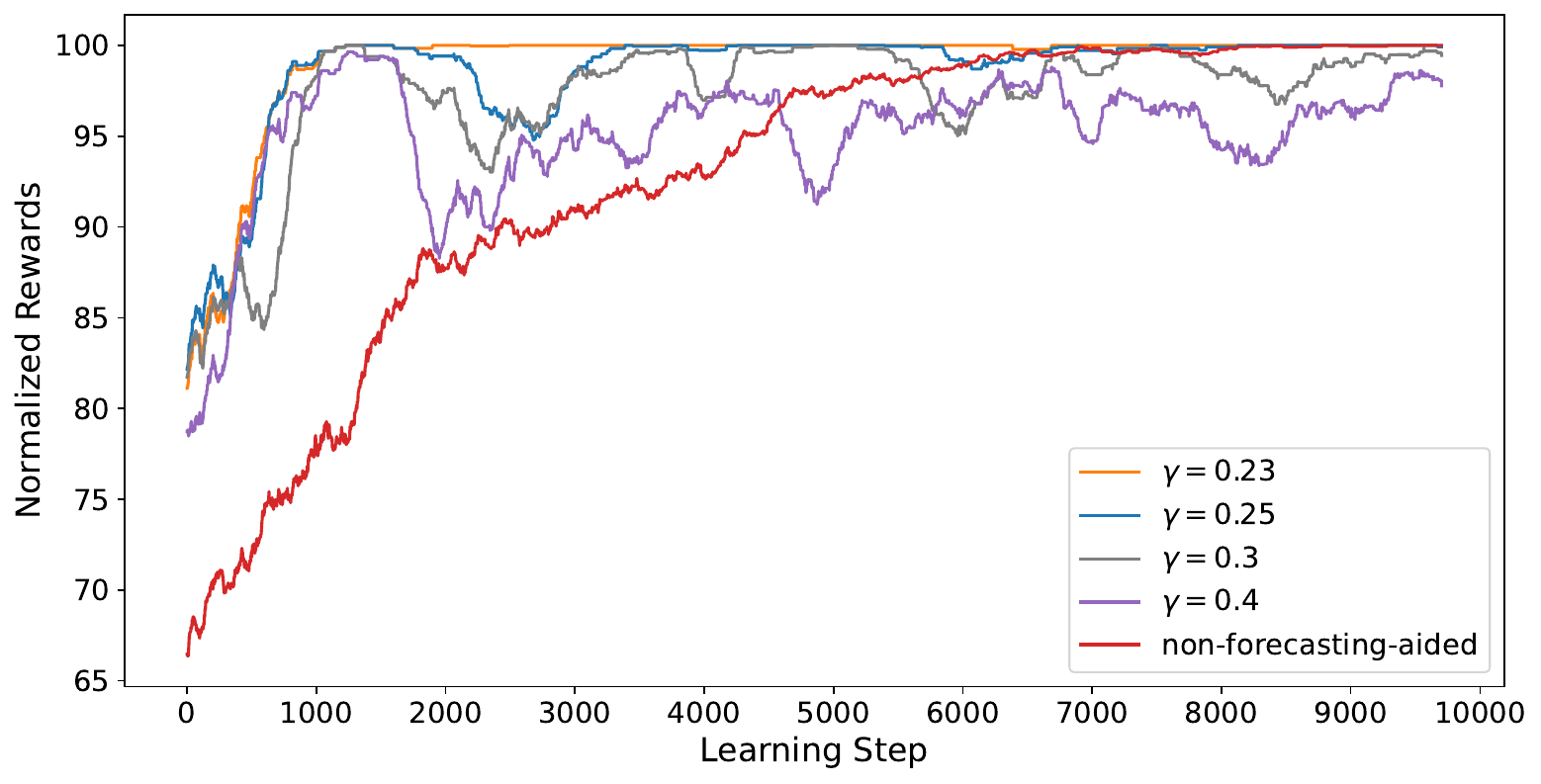}
\vspace{-1.5ex}
\setlength{\belowcaptionskip}{-15pt} 
\caption{Convergence performance of the proposed approach under different forecasting error models (traffic pattern 1).}
\label{fig:error}
\end{figure}

\paragraph{Forecast Error Effect}
\label{sec:error_effect}
We also examine the effect of the forecast error on the convergence performance. The proposed approach is robust against forecast errors that follow a Gaussian distribution with a standard deviation up to 0.25 in the case of traffic pattern 1 as shown in Fig. \ref{fig:error}. This is a significant error given that the range of the forecasted values is 1. Such robustness is attributed to accommodating potential forecasting errors in Algorithm \ref{alg:one} through divergence minimization instead of solely relying on the forecasted action. Nevertheless, when the standard deviation is higher than 0.25, the exploration becomes relatively unstable. DRL agents fail to converge in such scenarios. However, the overall reward is still kept in a relatively good range. Furthermore, the proposed algorithm still has relatively high initial reward values compared with the traditional DRL approach as seen in Fig. \ref{fig:error}.

\begin{figure}
\centering
\includegraphics[width=0.93\linewidth]{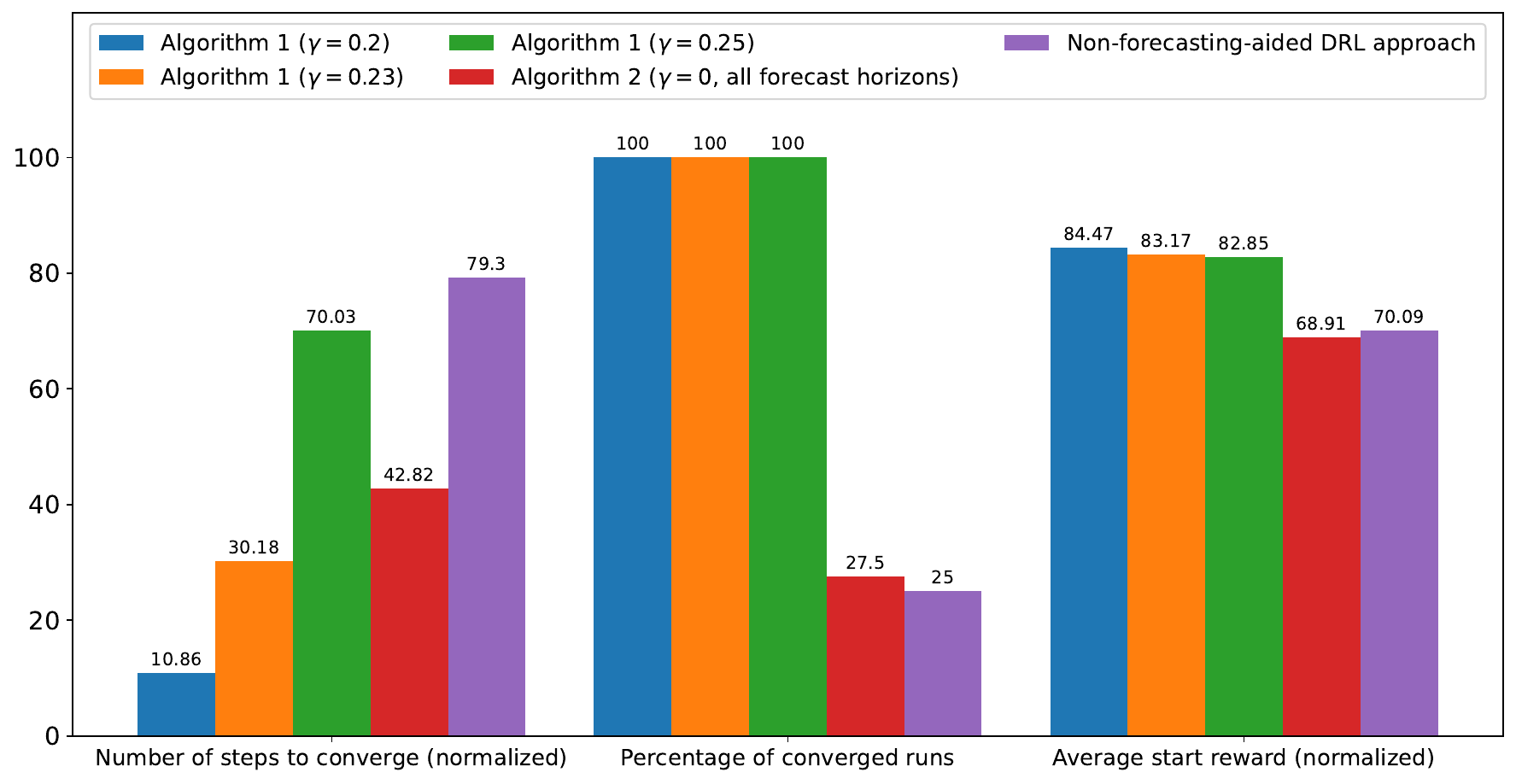}
\vspace{-1ex}
\setlength{\belowcaptionskip}{-15pt} 
\caption{Convergence performance averaged over multiple runs (the higher the better except for number of steps to converge).}
\label{fig:conv_perf}
\end{figure}

This observation is confirmed by the statistics compiled in Fig \ref{fig:conv_perf}. The proposed approach noticeably outperforms all the other DRL-based baselines and maintains the highest initial reward value on average, even when the error is high. Our approach also has the fastest convergence rate. Furthermore, 100\% of the conducted scenarios converge to the optimal resource allocation configuration given that the forecasting error's standard deviation is 0.25 or lower. Since forecasting models are imperfect, especially with new immersive services, this gives insights into the accepted error ranges. It also shows that forecasting models used in our approach do not have to be ideal. Finally, the forecasting-based state representation approach shows an inferior performance to the proposed approach in almost all the cases even when using perfect predictors.

\vspace{-0.4ex}

\section{Conclusion and Future Work}
\label{sec:conclusion}
In this paper, we conduct an exhaustive experiment to study the effect of forecasting on the convergence performance of DRL-based O-RAN slicing. We propose a \emph{forecasting-aided} DRL algorithm and an O-RAN deployment workflow that prove to remarkably enhance the convergence performance and to be robust against forecasting errors. We plan to investigate the possibility of building forecasting models that achieve the observed acceptable error ranges using the VR gaming data. We will also explore combining such models with other approaches such as constrained DRL and transfer learning as a promising step toward trustworthy DRL in O-RAN slicing.

\vspace{-0.4ex}

\bibliographystyle{IEEEtran}
\bibliography{references.bib}

\end{document}